\documentclass[10pt,twocolumn,twoside]{IEEEtran}
\usepackage{amsmath,amsopn,amssymb}
\usepackage{graphicx,xspace,color,soul}
\usepackage{epsfig,subfigure}
\usepackage{longtable,multirow}
\usepackage{array,float}
\usepackage{cite,citesort}
\usepackage{algorithm,algorithmicx,algpseudocode}
\usepackage{epstopdf}

\renewcommand{\vec}[1]{\mathbf{#1}}
\renewcommand{\algorithmicrequire}{\textbf{Input:}}
\renewcommand{\algorithmicensure}{\textbf{Output:}}

\usepackage{verbatim}

\newcolumntype{Y}{>{\centering\arraybackslash}X}

\begin{document}

\title{Predictive Generalized Graph Fourier Transform for Attribute Compression of Dynamic Point Clouds }
\author{
	Yiqun Xu,
	Wei Hu~\IEEEmembership{Member,~IEEE},
	Shanshe Wang~\IEEEmembership{Member,~IEEE},
	Xinfeng Zhang~\IEEEmembership{Member,~IEEE},
	Shiqi Wang~\IEEEmembership{Member,~IEEE},
	Siwei Ma~\IEEEmembership{Member,~IEEE},
	Zongming Guo~\IEEEmembership{Member,~IEEE},
	Wen Gao~\IEEEmembership{Fellow,~IEEE}

\begin{small}

\thanks{Corresponding authors: Wei Hu and Siwei Ma. 

Y. Xu is with Institute of Computing Technology, the University of Chinese Academy of Sciences, and also with the National Engineering Laboratory for Video Technology, Peking University,  (e-mail:yiqun.xu@vipl.ict.ac.cn). 

W. Hu and Z. Guo are with Wangxuan Institute of Computer Technology, Peking University (e-mail:forhuwei@pku.edu.cn; guozongming@pku.edu.cn).

S. Wang, S. Ma, and W. Gao are with the National Engineering Laboratory for Video Technology, Peking University (e-mail:sswang@pku.edu.cn; swma@pku.edu.cn; wgao@pku.edu.cn). 

X. Zhang is with the School of Computer and Control Engineering, University of the Chinese Academy of Sciences (email:xfzhang@ucas.ac.cn).

S. Wang is with the Department of Computer Science, City University of Hong Kong (e-mail:shiqwang@cityu.edu.hk). }

\end{small}
}
\maketitle
\vspace{0.1in}

\begin{abstract}
As 3D scanning devices and depth sensors advance, dynamic point clouds have attracted increasing attention as a format for 3D objects in motion, with applications in various fields such as immersive telepresence, navigation for autonomous driving and gaming. 
Nevertheless, the tremendous amount of data in dynamic point clouds significantly burden transmission and storage. 
To this end, we propose a complete compression framework for attributes of 3D dynamic point clouds, focusing on optimal inter-coding. 
Firstly, we derive the optimal inter-prediction and predictive transform coding assuming the Gaussian Markov Random Field model with respect to a spatio-temporal graph underlying the attributes of dynamic point clouds. 
The optimal predictive transform proves to be the Generalized Graph Fourier Transform in terms of spatio-temporal decorrelation. 
Secondly, we propose refined motion estimation via efficient registration prior to inter-prediction, which searches the temporal correspondence between adjacent frames of irregular point clouds. 
Finally, we present a complete framework based on the optimal inter-coding and our previously proposed intra-coding, where we determine the optimal coding mode from rate-distortion optimization with the proposed offline-trained $\lambda$-Q model. 
Experimental results show that we achieve around $17\%$ bit rate reduction on average over competitive dynamic point cloud compression methods.
\end{abstract}


\IEEEpeerreviewmaketitle

\section{Introduction}
\label{sec:intro}
\begin{figure}[t]
	\centerline{\psfig{figure=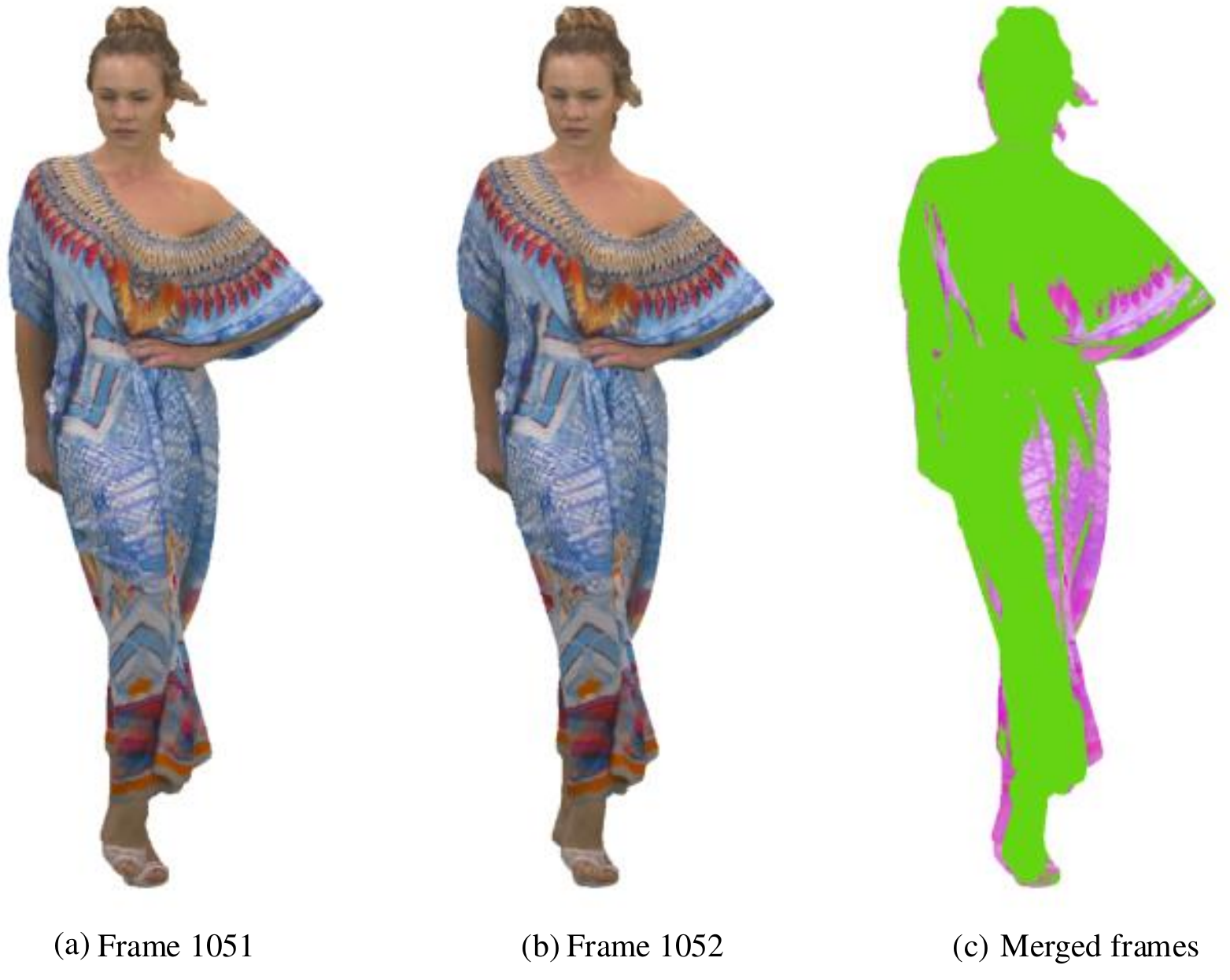,width=90mm} }
	\caption{Two frames in the dynamic point cloud sequence \textit{Longdress} \cite{Eugene2017dataset}, as well as the merged one to demonstrate the geometry difference between neighboring frames. In (c), frame 1051 is marked in purple while frame 1052 in green. }
	\label{fig1}
\end{figure}

With the development of depth sensing and 3D graphic technology, 3D \textit{dynamic} point clouds have attracted intensive attention in the representation for real scenes or 3D objects in motion. A dynamic point cloud consists of a sequence of static point clouds, each of which is composed of a set of discrete points, with 3D coordinates to represent the geometric information as shown in Fig.~\ref{fig1}. Attribute information is often associated with each point to enrich its functions, such as colors and normals. Due to the efficient representation, 3D dynamic point clouds have been widely applied in various fields, such as 3D immersive telepresence, navigation for autonomous vehicles, gaming, and animation \cite{Tulvan2016}.

Nevertheless, the large amount of data in 3D dynamic point clouds significantly increase the burden for transmission and storage, especially with multiple attributes on each point. Further, unlike images or videos, 3D points are acquired by non-uniformly sampling the surfaces of objects or scenes, which usually leads to \textit{irregular} point distribution. Also, the number of points in each frame often varies over time. Hence, it is quite challenging to represent dynamic point clouds compactly. In the face of such necessity and challenges of point cloud coding, the 3D Graphic Group (3DG) of Moving Picture Experts Group (MPEG) has started the standardization of point cloud compression (PCC)\footnote{https://mpeg.chiariglione.org/standards/mpeg-i/point-cloud-compression}.

Many approaches have been proposed to support efficient compression for \textit{static} point clouds, including geometry coding \cite{morell2014geometric,Schnabel2006,Radu2011,krivokuca2019volumetric} and attribute coding \cite{Florencio2014,xu2018cluster,de2016compression,gu20193d,chou2019volumetric},  as well as a few efforts for {\it dynamic} point clouds \cite{thanou2016graph,Khaled2017,anis2016compression,de2017motion,Ricardo2017mcic,mpeg2019verbal,pavez2017dynamic,pavez2018dynamic}.
However, existing dynamic point cloud coding methods focus more on the coding of geometry, but less on specialized inter-coding tools for attributes. In practice, attributes are critical in providing information for specific applications, \textit{e.g.}, colors and normals for visualization, as well as in rendering point clouds with high quality. Hence, we focus on the attribute coding of dynamic point clouds in this paper, assuming available geometry information at both the encoder and decoder.

In order to fully exploit intrinsic temporal correlations for compact representation, we propose \textit{optimal inter-prediction} and \textit{predictive transform} coding with refined motion estimation for attributes of dynamic point clouds.
Firstly, we represent dynamic point clouds on graphs, where each point is treated as a vertex and points are connected both spatially and temporally, leading to a spatio-temporal graph.
We then assume the Gaussian Markov Random Fields (GMRF) model \cite{rue2005gaussian} with respect to the spatio-temporal graph, and derive optimal inter-prediction and predictive transform coding for the prediction residual, which depends on the precision matrix in the GMRF model.
Instead of estimating the precision matrix statistically from multiple signal observations, we interpret the precision matrix by the \textit{generalized graph Laplacian matrix}\footnote{In spectral graph theory, the graph Laplacian is an algebraic representation of the connectivity and degree of a graph. There are different variants of Laplacian matrices, and the generalized graph Laplacian is one of them. The formal definition is in Section~\ref{sec:graph}.} in spectral graph theory \cite{Chung96,Shuman13}, which is able to be computed from a {\it single} signal observation.
The generalized graph Laplacian essentially encodes not only the spatial correlation within each frame, but also the temporal dependency across neighboring frames as the boundary condition of the current frame.
This leads to a generalized variant of Graph Fourier Transform (GFT), referred to as the generalized GFT (GGFT) \cite{hu15spl} as the optimal predictive transform.
The GGFT is an adaptive transform computed from the spatio-temporal graph underlying the signal, which proves to optimally decorrelate the dynamic point cloud both spatially and temporally under the assumed GMRF model.

Secondly, we propose refined motion estimation to address the challenge of searching the temporal correspondence between neighboring frames of irregular point clouds.
Specifically, we first segment each frame into clusters based on the spatial correlation of geometry, which serve as the processing unit for efficient representation.
For each target cluster in the current frame, in order to address the challenge of irregular sampling, we build a bounding cube around and register with a larger temporally collocated cube in the reference frame via Iterative Closest Point (ICP) \cite{chen1992object,besl1992method}.
This provides sufficient feasible searching space of corresponding points.
Then, we search the temporal correspondence based on point-to-point Euclidean distance.
This yields reference points for the subsequent inter-prediction and transform, which is able to accommodate complicated regions.

Finally, we design a complete framework of dynamic point cloud attribute compression with an offline-trained $\lambda$-Q model for rate-distortion optimization.
Specifically, the framework consists of two coding modes to fully exploit spatial-temporal dependencies, including our previous intra-coding \cite{xu2018cluster} as one mode and the proposed inter-coding as the other, both of which are based on GFTs.
We determine the optimal coding mode based on rate-distortion optimization to achieve the best trade-off between coding rates and attribute distortions.
Further, we establish a $\lambda$-Q model to derive the Lagrange multiplier \(\lambda\) off-line for best trade-off, according to the statistics of attributes in dynamic point clouds.
Taking color for instance, experimental results show that we outperform representative dynamic point cloud compression methods by around $17\%$ on average in bit rate reduction for the luminance component.
This validates the superiority of the proposed refined motion estimation, optimal inter-prediction and predictive GGFT coding for the spatial-temporal decorrelation of dynamic point clouds.

In summary, the main contributions of our work include:
\begin{itemize}

    \item We propose optimal inter-prediction and predictive transform coding assuming the GMRF model with respect to a spatio-temporal graph underlying the attributes of dynamic point clouds. We derive that the optimal predictive transform is the generalized GFT, which fully decorrelates the inter-prediction residual.

    \item We propose refined motion estimation via efficient registration prior to inter-prediction, which addresses the challenge of searching the temporal correspondence between neighboring frames of irregular point clouds.

    \item We present a complete framework of dynamic point cloud attribute compression, consisting of both intra-coding and inter-coding modes. The optimal coding mode is determined from rate-distortion optimization with the proposed offline-trained $\lambda$-Q model.
\end{itemize}


The remainder of this paper is organized as follows.
Section~\ref{sec:related} reviews related works on point cloud compression.
Section~\ref{sec:graph} introduces relevant concepts in spectral graph theory and the generalized GFT.
Next, we elaborate on the proposed optimal inter-prediction and predictive transform in Section~\ref{sec:prediction}.
Then we discuss the complete coding framework in Section~\ref{sec:framework}.
Experimental results and conclusions are presented in Section~\ref{sec:results} and Section~\ref{sec:conclude}, respectively.

\section{Related Work}
\label{sec:related}

Previous point cloud compression focuses more on the coding of geometry, such as the spinning tree \cite{gumhold2005predictive}, the kd-tree structure \cite{devillers2000geometric,ochotta2004compression}, and the octree \cite{Radu2011,Schnabel2006,kammerl2012real}, which organize unstructured point clouds into regular grids. Among them, the octree approach is widely adopted nowadays, which is a counterpart of quadtree in 2D images. Recently, the coding of attributes is drawing increasing attention. We discuss previous works in intra-coding and inter-coding of attributes in order.

\subsection{Intra-coding of point cloud attributes}
The intra-coding of point cloud attributes can be divided into two categories: 1) Codec-based methods, which leverage existing image/video codecs by projecting different viewpoints of point clouds onto depth maps and texture maps. 2) Geometry-based methods, which design transforms based on the geometry and perform transform coding on attributes.

In the first category, Mekuria \textit{et al.} employ JPEG \cite{wallace1992jpeg} to encode color information which is padded into a map via scanning in their proposed PCC-MP3DG method \cite{Mekuria2016}. Mammou \textit{et al.} \cite{Khaled2017} introduce the video-based point cloud compression (V-PCC), encoding the projected color and depth images via HEVC \cite{sullivan2012overview}, which improves the performance further. However, the scanning or projection process inevitably introduces losses in the structure and details of point clouds. Also, the characteristics of point clouds are not taken into consideration when encoding via existing image/video codecs.

In the second category, Huang \textit{et al.} \cite{Huang2008} explore the color attribute compression and verify that there is considerable redundancy in color representation within octree structure.
G-PCC \cite{3dg2019wd} is proposed as an open platform by MPEG-3DG, and Region-adaptive Hierarchical Transform (RAHT) \cite{de2016compression} is adopted as a tool of attribute coding.
Specifically, Queiroz \textit{et al.} compress the color information of point clouds via a hierarchical sub-band transform \cite{de2016compression} that resembles an adaptive variation of a Haar wavelet\cite{haar1910theorie}. They use an entropy coder to encode RAHT coefficients based on arithmetic coding (AC) and adaptive run-length Golomb-Rice encoding (RLGR) \cite{malvar2006adaptive} further, which is a much less complex entropy coder. Zhang \textit{et al.} \cite{Florencio2014} take one first step to represent point clouds on graphs and encode attributes via the GFT \cite{hammond2011wavelets} over each sub-cloud.
To further capture the spatial correlation,
our previous work proposes a Normal-Weighted GFT (NWGFT) \cite{xu2018cluster}. We first cluster the point cloud according to the point coordinate distribution, which makes each sub-cloud more correlated within. Then we construct the GFT by exploiting the similarity in normals of points, so that the correlation within each sub-cloud is further removed.

\subsection{Inter-coding of point cloud attributes}
Similarly, the inter-coding can be classified into two categories: 1) Codec-based methods \cite{Khaled2017}, which leverage existing video codecs (\textit{e.g.}, HEVC) but may lose details when projecting dynamic point clouds to videos. 2) Geometry-based methods, which exploit the temporal correlation in the 3D geometry. In \cite{pavez2017dynamic,pavez2018dynamic}, the polygon cloud is proposed to represent the geometry of dynamic point clouds. Moreover, the polygon cloud is compressed via transform coding of color and motion
residuals. Mekuria \textit{et al.} \cite{Mekuria2016} propose to register neighboring frames of point clouds via ICP, and encode the transformation matrix and other support information to reduce the bit rate. However, block-based replacement leads to severe distortion without any motion compensation tools during inter-coding.
In \cite{thanou2016graph}, Thanou \textit{et al.} represent a dynamic point cloud by a set of graphs and match features over them to perform motion estimation and predictive coding. However, the inter-coding residual is encoded based on the spatial correlation within each frame instead of the temporal correlation, leading to sub-optimal inter-coding. Queiroz \textit{et al.} propose motion-compensated intra-frame coder (MCIC) \cite{de2016compression,Ricardo2017mcic}, which is a novel motion-compensated approach based on the RAHT to encode dynamic voxelized point clouds.
However, due to the lack of effective inter-prediction residual coding, the performance of MCIC is limited to low bit rates.

\section{Background in Graph Fourier Transform}
\label{sec:graph}

\subsection{Graph, Graph Laplacian and Graph Fourier Transforms}
\label{subsec:related_graph}
We consider an undirected graph $ \mathcal{G}=\{\mathcal{V},\mathcal{E},\mathbf{W}\} $ composed of a vertex set $ \mathcal{V} $ of cardinality $|\mathcal{V}|=n$, an edge set $ \mathcal{E} $ connecting vertices, and a weighted \textit{adjacency matrix} $ \mathbf{W} $. $ \mathbf{W} $ is a real symmetric $ n \times n $ matrix, where $ w_{i,j} $ is the weight assigned to the edge $ (i,j) $ connecting vertices $ i $ and $ j $. We assume non-negative weights, \textit{i.e.}, $w_{i,j} \geq 0$.

The graph Laplacian matrix is then defined from the adjacency matrix. Among different variants of Laplacian matrices, the commonly deployed \textit{combinatorial graph Laplacian} \cite{Shen10,hu2015multiresolution} is defined as $\mathbf{L} := \mathbf{D} - \mathbf{W}$, where $ \mathbf{D} $ is the \textit{degree matrix}---a diagonal matrix with $ d_{i,i} = \sum_{j=1}^n w_{i,j} $. 

The combinatorial graph Laplacian is real and symmetric, which means it admits a complete set of orthonormal eigenvectors. The GFT basis $\mathbf{\Phi}$ is then the eigenvector set of the Laplacian matrix. For a given signal $\mathbf{f} \in \mathbb{R}^n$ defined on the vertices of graph $\mathcal{G}$ (\textit{i.e.}, \textit{graph signal}), the formal definition of its GFT is
\begin{equation}\label{formula:2}
\hat{\mathbf{f}} = \mathbf{\Phi}^{\top}\mathbf{f}.
\end{equation}
The inverse GFT follows as
\begin{equation}\label{formula:3}
\mathbf{f} = \mathbf{\Phi}\hat{\mathbf{f}}.
\end{equation}

The GFT is an adaptive linear transform, which has been shown to be superior in compressing a class of piecewise smooth signals, \textit{e.g.}, depth maps \cite{Shen2010,cheung2011depth,Hu12,hu2015multiresolution}. 

\subsection{The Generalized Graph Laplacian}
A generalized Laplacian (or \textit{discrete Schrodinger operator}) $\mathcal{L}$ is a symmetric matrix with non-positive off-diagonal entries \cite{biyikoglu2005nodal}. Defining a diagonal matrix $\mathbf{P} \in \mathbf{R}^{n \times n}$, one can rewrite a generalized Laplacian as   
\begin{equation}
    \mathcal{L} = \mathbf{L} + \mathbf{P},
    \label{eq:potential}
\end{equation}
where the diagonal entries of $\mathbf{P}$ can be viewed as a potential defined on the vertices, \textit{e.g.}, a boundary condition \cite{hu15spl}. It can also be written as
\begin{equation}
  \mathcal{L} = \mathbf{D}_g - \mathbf{W},  
  \label{eq5}
\end{equation}
where $\mathbf{D}_g$ is the degree matrix of the generalized Laplacian:
\begin{equation}
    \mathbf{D}_g = \mathbf{D} + \mathbf{P}.
\end{equation}


\subsection{The Generalized Graph Fourier Transforms}
The GGFT is firstly proposed in \cite{hu15spl} as an image transform, which is optimized for intra-prediction
residual in images. Following the definition of GFT, the basis of the GGFT is the eigenvector set $\mathbf{\Phi}_g$ of the generalized graph Laplacian $\mathcal{L}$. The GGFT is then defined as 
\begin{equation}
\hat{\mathbf{f}}_g = \mathbf{\Phi}_g^{\top}\mathbf{f}.
\label{eq:ggft}
\end{equation}
The inverse GGFT follows as
\begin{equation}
\label{formula:3}
\mathbf{f} = \mathbf{\Phi}_g\hat{\mathbf{f}}_g.
\end{equation}

In the generalized graph Laplacian in \cite{hu15spl}, the diagonal entries of vertices at image block boundaries are added with extra weights as a function of the expected inaccuracy of intra-prediction. It is analyzed in \cite{hu15spl} that the GGFT can be viewed as extensions of widely used transforms, namely, the discrete cosine transform (DCT) and the asymmetric discrete sine transform (ADST) \cite{han2012tip}.

\section{Optimal Inter-prediction and Transform}
\label{sec:prediction}
\begin{figure*}[t]
	\centerline{\psfig{figure=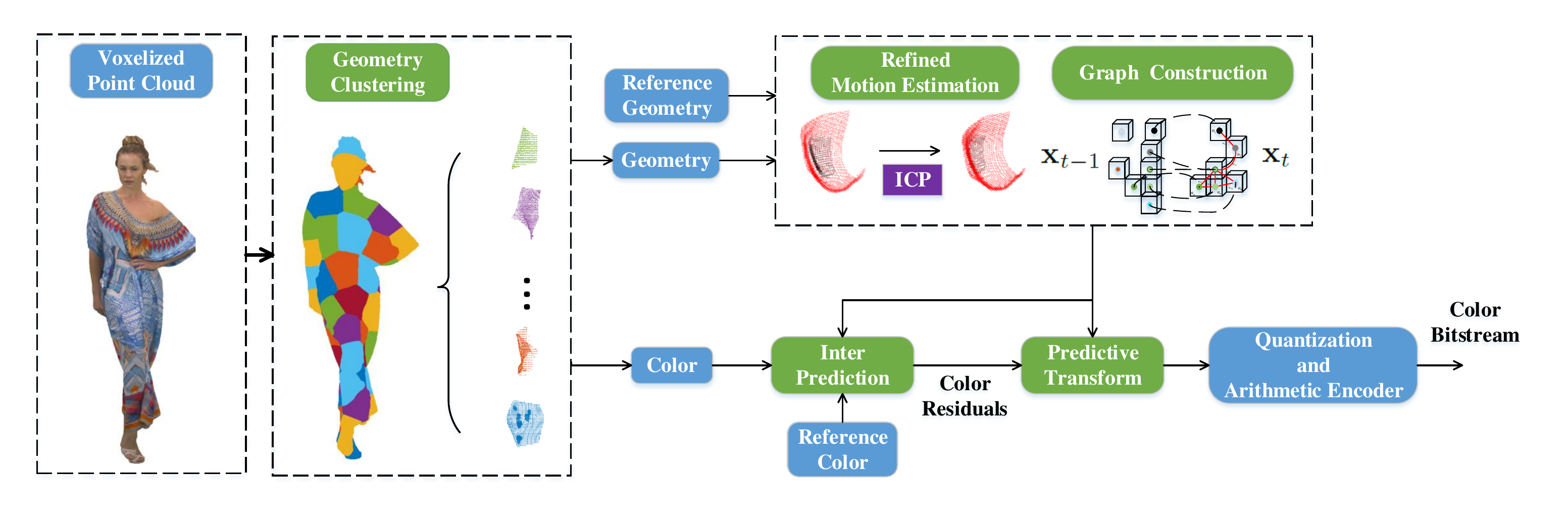,width=180mm} }
	\caption{The proposed inter encoder for attributes of dynamic point clouds.}
	\label{fig3}
\end{figure*}

In this section, we present the proposed inter-coding of dynamic point cloud coding, in particular the optimal inter-prediction and transform under GMRF modeling of a dynamic point cloud sequence.
As shown in Fig.~\ref{fig3}, the optimized inter-coding consists of four steps:
1) segment each frame of point cloud into clusters as the processing unit based on geometry;
2) search the temporal correspondence between each cluster in the current frame and a set of points in the previous frame, \textit{i.e.}, refined motion estimation;
3) construct a spatio-temporal graph over each cluster in the current frame and compute its generalized Laplacian matrix;
4) perform optimal inter-prediction and predictive transform.
We discuss the four steps in order as follows.

\subsection{Geometry Clustering}
\label{subsec:clustering}
Given an input point cloud sequence $\mathbf{P}=\{\mathbf{P}_1,...,\mathbf{P}_l\}$, where $l$ is the number of frames in $\mathbf{P}$, we consider two neighboring frames $\mathbf{P}_{t-1}$ and $\mathbf{P}_{t}$ $ (2 \leq t \leq l)$.
To capture the local characteristics of each frame,
we partition the target point cloud frame $\mathbf P_t$ into small clusters.
Instead of uniform spatial partition of point clouds~\cite{Florencio2014} which would create many isolated sub-clouds if the point cloud is sparse, we employ $K$-means clustering~\cite{everingham2005automated} based on geometry, assuming availability of geometry at both the encoder and decoder.
The point cloud is divided into $K$ clusters and the $k$-th cluster contains $n_k$ points.
To balance the coding performance and computational complexity, we set the {\it average} number of points in each cluster to $600$.
Moreover, $K$-means clustering keeps the geometric continuity of sub-clouds in each cluster.
The subsequent steps are performed on cluster-basis.


\subsection{Refined Motion Estimation}

\begin{figure}[t]
	\centerline{\psfig{figure=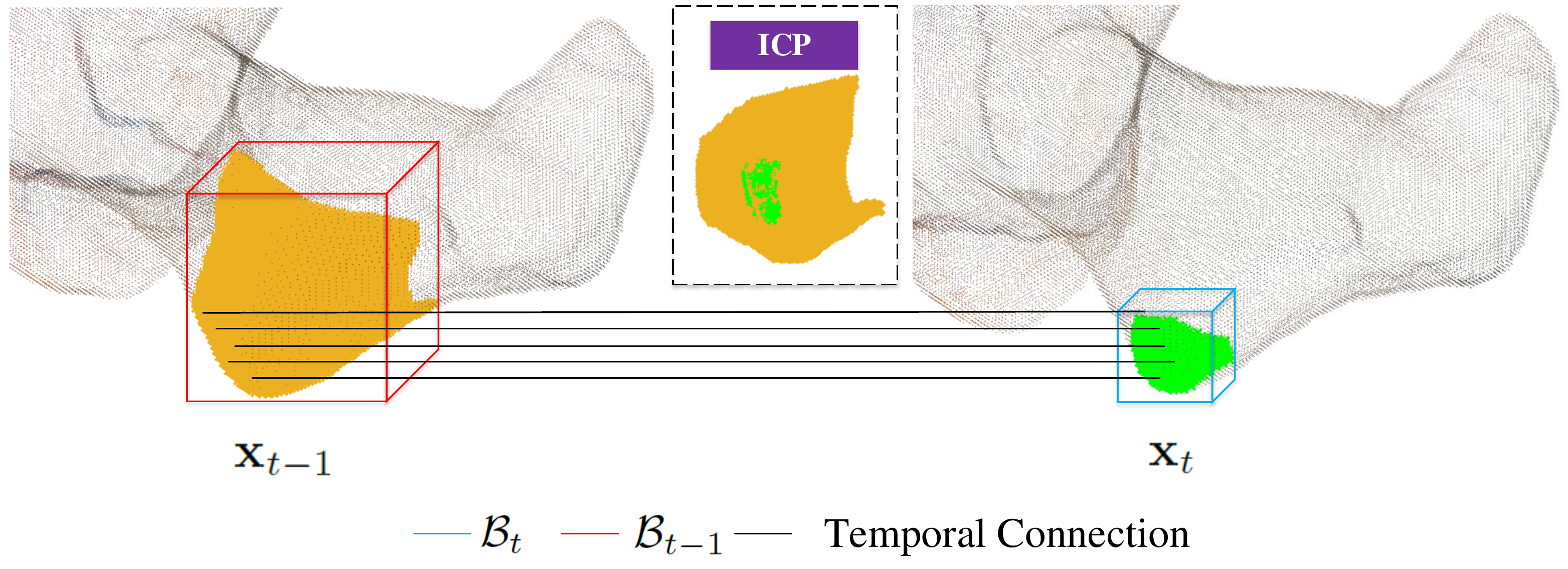,width=90mm} }
	\caption{Illustration of the proposed refined motion estimation. The green patch is the target cluster and the yellow one is the reference set.}
	\label{fig:relation}
\end{figure}

In order to efficiently exploit the temporal correlation between the input frame $\mathbf{P}_{t}$ and its previous frame $\mathbf{P}_{t-1}$, we propose to register each target cluster $\mathbf{x}_t$ in $\mathbf{P}_{t}$ with the previous frame $\mathbf{P}_{t-1}$ via ICP and then find point-to-point correspondence in registered sets.

As demonstrated in Fig.~\ref{fig:relation}, to reduce the registration complexity, we first form a bounding box $\mathcal{B}_t$ around $\mathbf{x}_t$.
Then, we set a bounding box $\tilde{\mathcal{B}}_{t-1}$ in $\mathbf{P}_{t-1}$, which is collocated with $\mathcal{B}_t$ and expanded with a certain percentage $\delta$ ($\delta=300\%$ in our experiments).
Next, we register the points in $\tilde{\mathcal{B}}_{t-1}$ to align with $\mathbf{x}_t$ via ICP, leading to a registered point set $\mathcal{B}_{t-1}$ that will be employed to find reference points in $\mathbf{P}_{t-1}$.
This serves as a {\it refinement} step for motion estimation.

Further, we acquire point-to-point correspondence between points of the target cluster $\mathbf{x}_t$ in $\mathbf{P}_t$ and points in the reference bounding box $\mathcal{B}_{t-1}$ in $\mathbf{P}_{t-1}$.
Specifically, we globally search for temporally corresponding points in $\mathcal{B}_{t-1} $ based on the Euclidean distance in geometric coordinates.
The temporal correspondence of each point $\mathbf{p}_i$ in $\mathbf{x}_t$ is its nearest point $\mathbf{p}_j$ in $\mathcal{B}_{t-1}$ in terms of Euclidean distance among all the points.
The resulting $\{\mathbf{p}_j\}_{j=1}^{n_k}$ forms the corresponding set (unnecessarily a connected cluster) in $\mathbf{P}_{t-1}$, denoted as $\mathbf{x}_{t-1}$.
As such, we acquire the temporal correspondence between the reference frame $\mathbf{P}_{t-1}$ and the current frame $\mathbf{P}_t$.

Note that, while we accommodate neighboring frames $\mathbf{P}_{t-1}$ and $\mathbf{P}_{t}$ with different number of points, the $k$-th cluster $\mathbf{x}_t$ in $\mathbf{P}_{t}$ and its corresponding set $\mathbf{x}_{t-1}$ in $\mathbf{P}_{t-1}$ contain the same number of points $n_k$ by the proposed refined motion estimation.

\subsection{Proposed Spatio-temporal Graph Construction}
\label{subsec:graph_construction}

We construct a spatio-temporal graph over $\mathbf{x}_t$ and $\mathbf{x}_{t-1} $ to encode spatio-temporal correlations, which prepares the ground for the subsequent inter-prediction and predictive transform.
Specifically, we treat each point as a vertex in the graph, and build connections including \textit{spatial connectivities} within $\mathbf{x}_t$ and \textit{temporal connectivities} between corresponding points of $\mathbf{x}_t$ and $\mathbf{x}_{t-1}$.
We discuss the spatial and temporal connectivities as follows.

\subsubsection{Spatial graph connectivities}

We construct spatial connectivities in $\mathbf{x}_t$ based on global and local features as in our previous work \cite{xu2018cluster}.
Specifically, we build an $\epsilon$-neighborhood graph by calculating the Euclidean distance among 3D points globally to decide neighbors of each point within a radius $\epsilon$.
We then estimate the normal vector at each point from the local surface constructed over the current point and its $k$-nearest-neighbors, which serves as the local feature to compute edge weights in a Gaussian kernel.
More details will be described in Sec.~\ref{subsec:define_d}.

\subsubsection{Temporal graph connectivities}

We build temporal connections based on point-to-point correspondence acquired from the refined motion estimation.
Specifically, we connect each point in the target cluster $\mathbf{x}_t$ of $\mathbf{P}_t$ to its corresponding point in $\mathbf{x}_{t-1}$ of $\mathbf{P}_{t-1}$.
While it is possible to set weights of temporal edges as a Gaussian kernel of the distance between 3D coordinates of corresponding points, we assign all the temporal edge weights as $1$ assuming maximum temporal correlation.
This is because the temporal correspondence is accurate in general based on the refined motion estimation.
Though the searched temporal correspondence may differ in geometric coordinates due to the motion or sampling, they essentially represent the same region of a 3D object, thus we assume the maximum temporal correlation and set temporal edge weights as $1$.

\subsection{Optimal Inter-Prediction and Transform under GMRF}
We propose optimal inter-prediction and predictive transform based on the constructed spatio-temporal graph.
In particular, we derive optimal prediction and transform of point clouds statistically under the GMRF model, which plays a critical role in describing the spatial and temporal dynamics of nature and real systems \cite{rue2005gaussian}.

\subsubsection{Preliminaries in Gaussian Markov Random Field}

The formal definition of a GMRF is as follows.

\textbf{Definition} A random vector $\vec{x}=(x_1,...,x_n)^{\top}$ is called a GMRF with respect to the graph $\mathcal{G}=\{\mathcal{V}=\{1,...,n\},\mathcal{E}\}$ with mean $\vec{\mu}$ and a precision matrix $\mathbf{Q} \succ 0$ (positive definite), if and only if its density has the form
\begin{equation}
    p(\vec{x}) = (2\pi)^{-\frac{n}{2}}|\mathbf{Q}|^{\frac{1}{2}} \text{exp}(-\frac{1}{2}(\vec{x}-\vec{\mu})^{\top}\mathbf{Q}(\vec{x}-\vec{\mu})),
    \label{eq:GMRF}
\end{equation}
and
\begin{equation}
    q_{i,j} \neq 0 \Longleftrightarrow \{i,j\} \in \mathcal{E},~\forall i \neq j
    \label{eq:GMRF_graph}
\end{equation}

This definition infers that a signal $\vec{x}$ modeled by a GMRF follows a multivariate Gaussian distribution, with the mean vector $\vec{\mu}$ and the covariance matrix $\mathbf{\Sigma}$---the inverse of $\mathbf{Q}$.
The precision matrix is deployed in the definition of GMRF for its conditional interpretations:
\begin{align}
    &E(x_i|\vec{x}_{-i})=\mu_i-\frac{1}{q_{i,i}}\sum\limits_{j:j \sim i} q_{i,j}(x_j-\mu_j),
    \label{eq:conditional_mean} \\
    &Prec(x_i|\vec{x}_{-i})=q_{i,i},
    \label{eq:conditional_precision}
\end{align}
where $\vec{x}_{-i}$ denotes all elements in $\vec{x}$ except $x_i$, and $j:j \sim i$ represents all nodes $j$ that are neighbors of $i$ in the graph. \eqref{eq:conditional_mean} and \eqref{eq:conditional_precision} interpret the conditional expectation and precision of $x_i$ given all other elements based on parameters $(\vec{\mu},\mathbf{Q})$ of the GMRF, which will be leveraged to derive an optimal inter-prediction and predictive transform.

\begin{figure*}[t]
	\centerline{\psfig{figure=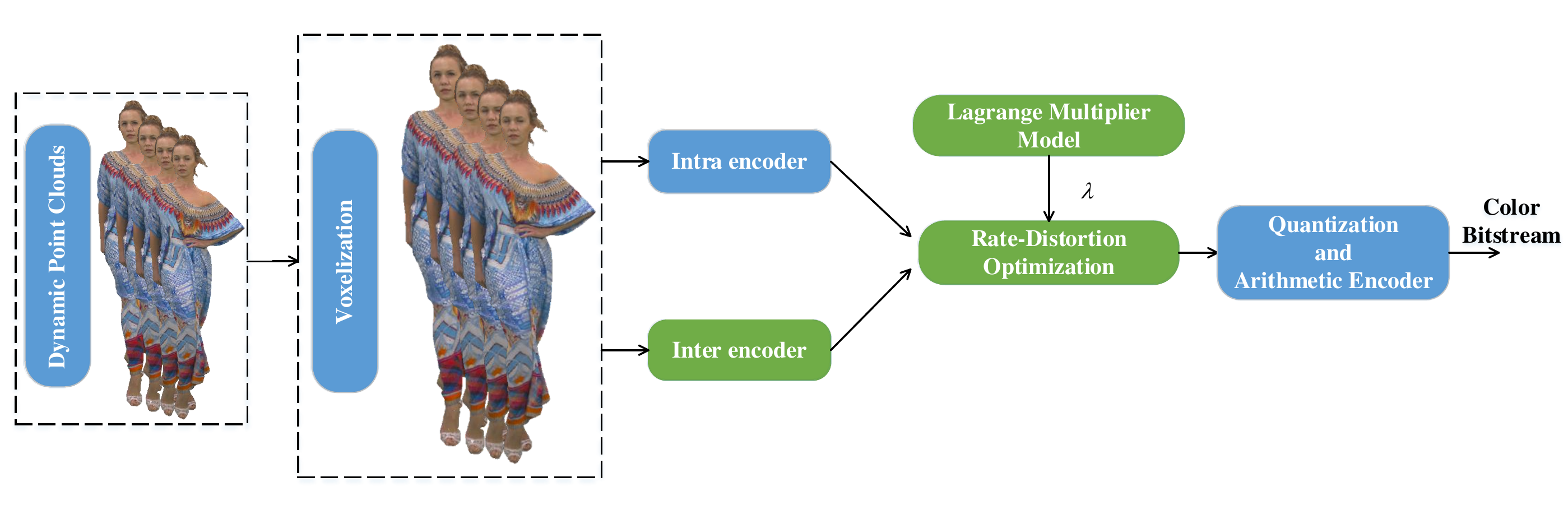,width=170mm} }
	\caption{The proposed coding framework for attributes of 3D dynamic point clouds.}
	\label{fig2}
\end{figure*}

\subsubsection{Optimal prediction and transform}

We assume the attributes of $\mathbf{x}_{t-1}$ and $\mathbf{x}_t$ follow the GMRF model, with mean $\vec{\mu}_{\vec{x}_t}$ and $\vec{\mu}_{\vec{x}_{t-1}}$, and the precision matrix $\mathbf{Q}_c$ and  $\mathbf{Q}_{\text{ref}}$ respectively.

Given the reference set $\mathbf{x}_{t-1}$, the inter-prediction problem is essentially predicting $\mathbf{x}_{t}$ from $\mathbf{x}_{t-1}$.
The \textit{optimal inter-prediction} is the conditional expectation $\vec{\mu}_{\vec{x}_t | \vec{x}_{t-1}}$ of $\mathbf{x}_{t}$ given $\mathbf{x}_{t-1}$ under the GMRF model.
Any other predictor will yield non-zero expected prediction error.
For the resulting prediction residuals, the \textit{optimal predictive transform} basis is the eigenvector set of the precision matrix $\mathbf{Q}_{{\mathbf{x}_t}|{\mathbf{x}_{t-1}}}$, {\it i.e.}, the Karhunen-Loeve Transform (KLT) \cite{vranic2001tools}, to optimally decorrelate $\mathbf{x}_{t}$ under the GMRF.

In particular, we group $\mathbf{x}_{t-1}$ and $\mathbf{x}_t$ into $\mathbf{x} = [\mathbf{x}_t^{\top}, \mathbf{x}_{t-1}^{\top}]^{\top}$, with mean $\mu \in \mathbb{R}^{2n \times 1} $ and precision matrix $\mathbf{Q} \in \mathbb{R}^{2n \times 2n}$ as the parameters\footnote{We denote the number of points in $\mathbf{x}_{t-1}$ and $\mathbf{x}_t$ by $n$ instead of $n_k$ for simplicity since here, though various clusters may contain different number of points.}. Then we partition the precision matrix $\mathbf{Q}$ as
\begin{equation}
\mathbf{Q} = \left( {\begin{array}{*{20}{c}}
{{\mathbf{Q}_{t,t}}}&{{\mathbf{Q}_{t,t-1}}}\\
{\mathbf{Q}_{t,t-1}^{\top}}&{{\mathbf{Q}_{t-1,t-1}}}
\end{array}} \right),
\label{eq:Q}
\end{equation}
where $\mathbf{Q}_{t,t} \in \mathbb{R}^{n \times n}$, $\mathbf{Q}_{t,t-1} \in \mathbb{R}^{n \times n}$ and $\mathbf{Q}_{t-1,t-1} \in \mathbb{R}^{n \times n}$ are the submatrices.

As mentioned in \cite{rue2005gaussian}, ${\mathbf{x}_t}|{\mathbf{x}_{t-1}}$ is also a GMRF with mean $\mu_{{\mathbf{x}_t}|{\mathbf{x}_{t-1}}}$ and precision matrix $\mathbf{Q}_{{\mathbf{x}_t}|{\mathbf{x}_{t-1}}}$.
According to \eqref{eq:conditional_mean} and \eqref{eq:conditional_precision}, we have
\begin{align}
    &\vec{\mu}_{\vec{x}_t | \vec{x}_{t-1}}=
    \vec{\mu}_{\vec{x}_t} - \mathbf{Q}_{t,t}^{-1}\mathbf{Q}_{t,t-1}(\vec{x}_{t-1}-\vec{\mu}_{\vec{x}_{t-1}}),
    \label{eq:prediction1}\\
    &\mathbf{Q}_{\vec{x}_t | \vec{x}_{t-1}}=\mathbf{Q}_{t,t}.
    \label{eq:transform1}
\end{align}

Next, we discuss the specific form of $\mathbf{Q}_{t,t-1}$ and $\mathbf{Q}_{t,t}$ based on our spatio-temporal graph construction.
According to (\ref{eq:GMRF_graph}), $\mathbf{Q}_{t,t-1}$ represents the temporal connectivity between $\mathbf{x}_t$ and $\mathbf{x}_{t-1}$.
In particular, the precision matrix $\mathbf{Q}$ in \textit{general GMRF} can be interpreted by a \textit{generalized} graph Laplacian $\mathcal{L}$~\cite{pavez2016generalized,egilmez2017graph,zhang2015graph}, \textit{i.e.},
 \begin{equation}
     \mathbf{Q}=\mathcal{L}.
     \label{eq36}
 \end{equation}
This will be further validated experimentally in Section VI-B. 

As discussed earlier, we connect each pair of temporally corresponding points in $\mathbf{x}_t$ and $\mathbf{x}_{t-1}$ and assign each edge weight $w_{i,j}$ as $1$. From the definition of the graph Laplacian in Section~\ref{sec:graph}, we have $l_{i,j}=-w_{i,j}=-1$. Together with \eqref{eq36}, we have
\begin{equation}
    \mathbf{Q}_{t,t-1} = -\mathbf{I},
    \label{eq:Q_t,t-1}
\end{equation}
where $\mathbf{I} \in \mathbb{R}^{n \times n}$ is an identity matrix, and
\begin{equation}
    \mathbf{Q}_{t,t}=\mathcal{L}_{t,t} = \mathbf{L}_t + \mathbf{I},
    \label{eq26}
\end{equation}
where $\mathbf{L}_t$ encodes the spatial connectivities of $\vec{x}_t$ and $\mathbf{I}$ encodes the temporal connectivities to $\vec{x}_{t-1}$.
Here $\mathbf{I}$ corresponds to $\mathbf{P}$ in \eqref{eq:potential}, which can be viewed as the temporal condition of each cluster in the current frame.



Substituting \eqref{eq:Q_t,t-1} and \eqref{eq26} into \eqref{eq:prediction1}, we have the optimal prediction as
\begin{equation}
    \vec{\mu}_{\vec{x}_t | \vec{x}_{t-1}}=
    \vec{\mu}_{\vec{x}_t} + (\mathbf{L}_t + \mathbf{I})^{-1}(\vec{x}_{t-1}-\vec{\mu}_{\vec{x}_{t-1}}).
\end{equation}

Further, assuming zero-mean for both $\vec{x}_t$ and $\vec{x}_{t-1}$, \textit{i.e.}, $\vec{\mu}_{\vec{x}_t}=\vec{\mu}_{\vec{x}_{t-1}}=\vec{0}$, we have
\begin{equation}
    \vec{\mu}_{\vec{x}_t | \vec{x}_{t-1}}=
    (\mathbf{L}_t + \mathbf{I})^{-1}\vec{x}_{t-1}.
    \label{eq29}
\end{equation}
This is the optimal inter-prediction we derive and employ.
As $\mathbf{L}_t$ is a high-pass filter, the inter-prediction of $\vec{x}_t$ is essentially a {\it low-pass filtered} version of the reference set $\vec{x}_{t-1}$, instead of directly copying from $\vec{x}_{t-1}$.

Substituting \eqref{eq26} into \eqref{eq:transform1}, the optimal predictive transform basis is the eigenvector set of
\begin{equation}
    \mathbf{Q}_{\vec{x}_t | \vec{x}_{t-1}} = \mathcal{L}_{\vec{x}_t | \vec{x}_{t-1}}=\mathbf{L}_t + \mathbf{I}.
    \label{eq30}
\end{equation}
As $\mathcal{L}_{\vec{x}_t | \vec{x}_{t-1}}$ is a generalized Laplacian, the resulting transform is GGFT \cite{hu15spl}.
By incorporating temporal dependencies between adjacent frames, the derived GGFT is optimal in terms of \textit{full decorrelation} as an interpretation of the optimal KLT.
Note that, we may employ the fast GFT algorithm in \cite{le2017approximate} to accelerate the eigen-decomposition of the graph Laplacian for the GGFT, reducing the complexity from $\mathcal{O}(n^3)$ to $\mathcal{O}(n \log n)$ while exhibiting constant error for growing $n$.
We then perform the GGFT as in \eqref{eq:ggft}.
The resulting transform coefficients are quantized, entropy encoded and transmitted to the decoder.

\section{Proposed Dynamic Point Cloud Coding Framework}
\label{sec:framework}
Based on the optimal inter-prediction and predictive transform, we present a complete coding framework for attributes of 3D dynamic point clouds, including the proposed inter-coding and our previously proposed intra-coding \cite{xu2018cluster} as shown in Fig.~\ref{fig2}.
Further, in a predictive frame, we design mode decision to choose between the {\it inter-coding mode} and {\it intra-coding mode} so as to achieve better Rate-Distortion performance for each cluster.
Note that, we assume the geometry information of point clouds is {\it available} at both the encoder and decoder.
The geometry could be uncompressed as is, or compressed in a lossy or lossless manner with existing methods.


\subsection{Preprocessing: Voxelization}
Different point clouds have various scales of \textit{size} and \textit{precision}, which inevitably affects the inter/intra-coding process as we construct graphs based on local and global features of geometry.
For the sake of generality, we preprocess the input point cloud via \textit{voxelization}, which maps all the points into bins of dimension $N \times N \times N $ and thus leads to point clouds with unified scale in coordinates ($N$ is set to 4096 in our experiments).
Specifically, a bin or a voxel is regarded occupied, if it contains at least one point, otherwise it is unoccupied.
Then the original point cloud is voxelized by quantizing the point position to a regular lattice, and represented by the voxelized point cloud during our coding process.
The attribute (\textit{e.g.}, color) of each voxel is calculated as the average attribute value of all the points within the voxel.
At the decoder, we perform \textit{devoxelization} on the decoded point clouds so as to comply with the objective evaluation metric of MPEG and keep the geometric scale of point clouds unchanged.


\begin{figure*}[t]
	\centerline{\psfig{figure=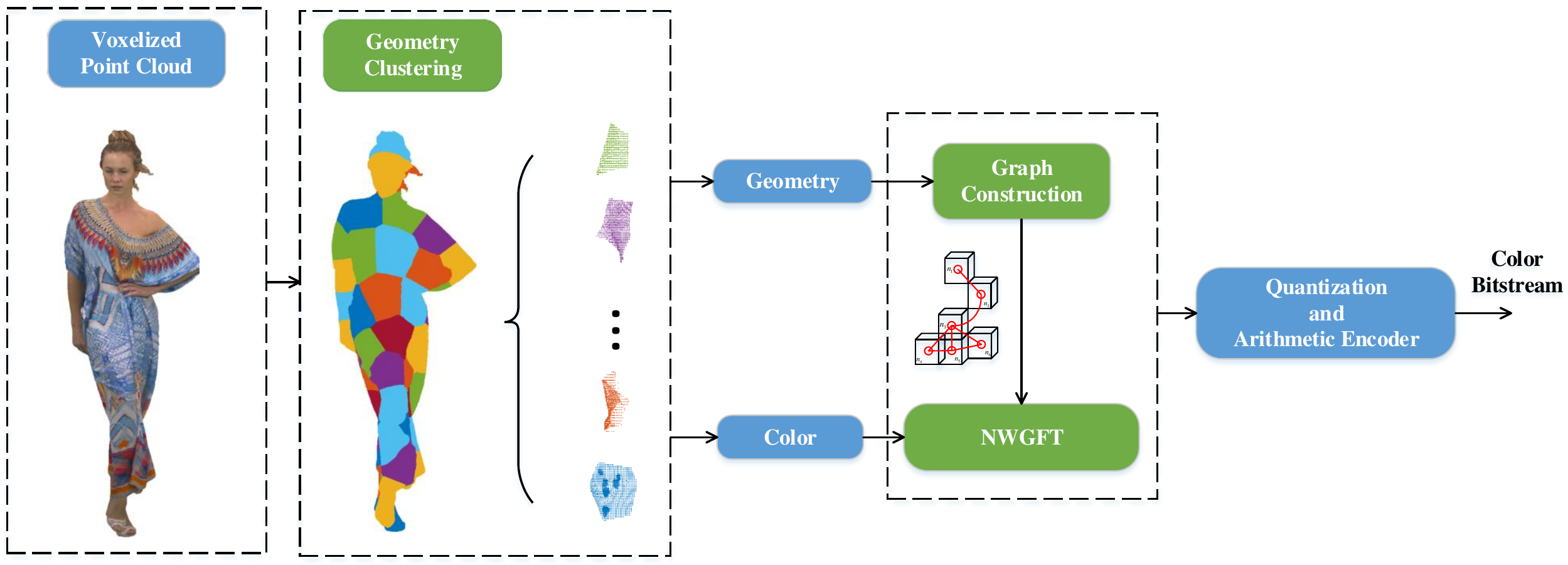,width=170mm} }
	\caption{The proposed intra encoder for attributes of dynamic point clouds.}
	\label{fig4}
\end{figure*}

\subsection{Intra-Coding with Normal-Weighted GFT}
\label{subsec:define_d}
Subsequent to voxelization, we perform geometry clustering as discussed in Section~\ref{subsec:clustering}, where each cluster serves as the processing unit of intra-coding as in inter-coding.
As illustrated in Fig.~\ref{fig4}, we adopt our previous algorithm Normal-Weighted Graph Fourier Transform (NWGFT) in \cite{xu2018cluster} for intra-coding.
The key idea is to capture structural similarities by a Gaussian kernel of normals as edge weights, from which the GFT is computed for compact representation.

In NWGFT, we propose a novel edge weight allocation method for graph construction by exploiting the similarity of normal vectors in a local space to further remove the correlation of each cluster.
Specifically, we first build an $\epsilon$-neighborhood graph, which connects all points whose pairwise distances are smaller than $\epsilon$. The choice of $\epsilon$ is dependent on the density of the entire point cloud, which we will detail in the experimental setup.
Next, we compute the normal vector $\vec{n}_i$ of a $k$-nearest-neighbor local space around each point $\vec{p}_i$. The normal is estimated by decomposing the covariance matrix \cite{kpfrs1901lines}, which serves as a local feature.
The edge weight $w_{i,j}$ between $\vec{p}_i$ and $\vec{p}_j$ is then assigned as:
\begin{equation}\label{formula:edgeweight}
    {w_{i,j}} = {e^{ - {{\left( {\frac{{\sin \theta_{i,j} }}{\sigma }} \right)}^2}}},
\end{equation}
where $\theta_{i,j}$ is the angle between two normal vectors $\vec{n}_i$ and $\vec{n}_j$, and $ \sigma $ is a parameter in the Gaussian kernel.
In our experiments, $k$ and $\sigma^2$ are empirically set to $15$ and $0.4$, respectively.
The proposed edge weight function defined in (\ref{formula:edgeweight}) is more robust than commonly used Euclidean distance between coordinates \cite{Florencio2014}, by considering features of each point and its neighborhood via normals.

Based on the above graph construction, we compute a combinatorial graph Laplacian to acquire the GFT basis for each cluster.
Similar to the inter-coding, we may deploy the fast GFT \cite{le2017approximate} to reduce the computation complexity of the GFT.
The resulting transform coefficients are quantized, entropy encoded and transmitted to the decoder.

\subsection{Proposed Coding Mode Decision}
In the proposed coding system, each cluster in a predictive frame has two coding modes: the inter-coding mode as elaborated in Section~\ref{sec:prediction} and the intra-coding mode discussed in Section~\ref{subsec:define_d}.
We design a decision scheme to choose the best coding mode.
Given a target bit rate, the goal of the attribute compression of dynamic point clouds is to convey the attribute information with minimum possible distortion.
In order to achieve the best rate-distortion performance, a coding mode should be determined from the intra and inter modes to reach the best balance between rate and distortion.
This can be cast into the classical rate-distortion framework, which is formulated as pursuing the best quality under the limitation of a given target rate \cite{Sullivan1998,zhang2017just} for each cluster,
\begin{equation}
\begin{array}{*{20}{c}}
 {\mathop {\min }\limits_{\{ {m_i}\} }  {{D}\left( {{m_i}} \right)} } \quad\quad \;\;\;\;\;\;\;\;\;\;\;\;\;\;\;\;\;\;\;\;\;\;\;\;\;\;\;\; \\
 {\text{subject to} \quad  {{R}\left( {{m_i}} \right)}  \le {R_T}}, i=1,2,
\end{array}
\end{equation}
where ${R_T}$ indicates the target bits, $m_i$ denotes the intra-coding mode ($m_1$) or the inter-coding mode ($m_2$), ${D}({m_i})$ and ${R}({m_i})$ denote the distortion and coding bits respectively when employing the $i$-th coding mode $m_i$.
This constrained problem can be converted into the unconstrained Lagrangian rate-distortion optimization problem as
\begin{equation}\label{equ:rdequ}
\mathop {\min }\limits_{\{ {m_i}\} } {J}\left( {{m_i}} \right)  \quad \text{where}  \quad {J}\left( {{m_i}} \right) = {D}\left( {{m_i}} \right) + {\lambda}\cdot{R}\left( {{m_i}} \right), i=1,2,
\end{equation}
where ${J}\left( {{m_i}} \right)$ is the rate-distortion cost of mode $m_i$, and $\lambda$ is the Lagrange multiplier which controls the trade-off between rate and distortion.
In particular, \(\lambda\) is determined by the quantization factor \(Q\), leading to a $\lambda$-Q model.
Given \(\lambda\), the optimal mode with the lowest rate-distortion cost can be identified.
We discuss the rate-distortion calculation and $\lambda$-Q model in detail below.

\subsubsection{Rate and Distortion Calculation}
We calculate the coding rate of mode $m_i$ in terms of \textit{bytes per voxel} after encoding the attribute of a frame.
Taking color for instance, the distortion $D(m_i)$ is obtained using the average of Mean Squared Error (MSE) in the YUV space:
\begin{equation}\label{formula:7}
D({m_i}) = \frac{{MS{E_Y}({m_i}) + MS{E_U}({m_i}) + MS{E_V}({m_i})}}{3},
\end{equation}
where \(MS{E_Y}(m_i)\), \(MS{E_U}(m_i)\) and \(MS{E_V}(m_i)\) denote the distortion between the original and reconstructed frames in components Y, U and V, respectively.

 \begin{figure}[t]
 	\centerline{\psfig{figure=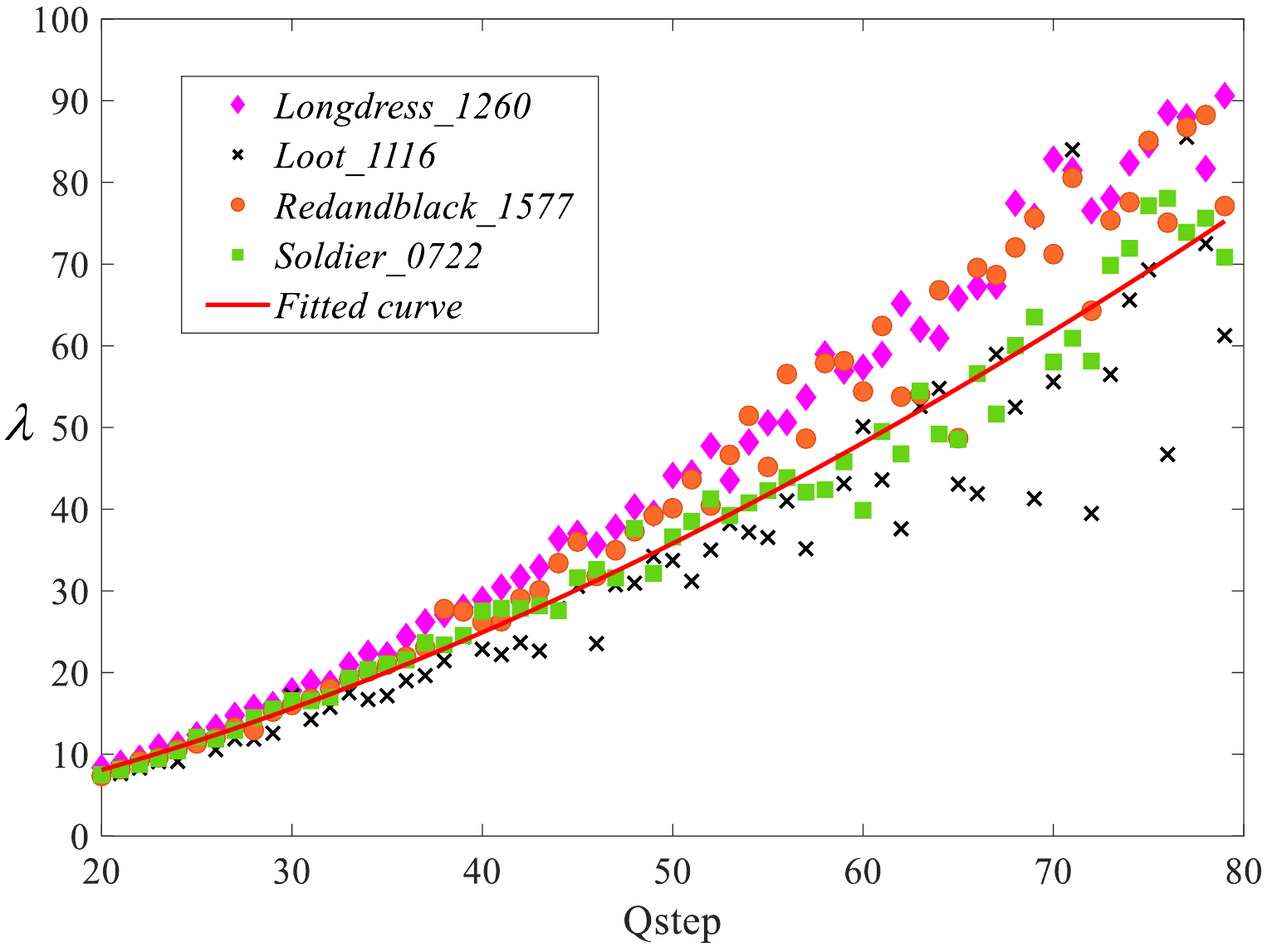,width=85mm} }
 	\caption{ The relationship between Lagrange multiplier $\lambda$ and Qstep for different dynamic point clouds.}
 	\label{fig55}
 \end{figure}

\subsubsection{$\lambda$-Q model}
In traditional image and video compression, the Lagrange multiplier $\lambda$ as a function of \(Q\)  can be off-line trained from the statistics of images or videos. Analogously, in the context of point cloud compression, we learn an appropriate $\lambda$-Q model for efficient rate-distortion optimization. By setting the derivative of \(J\) in \eqref{equ:rdequ} with respect to the quality factor $Q$ to 0,
\begin{equation}\label{formula:4}
\frac{{\partial {J} }}{{\partial Q} } = \frac{{\partial {D} }}{{\partial Q} } + {\lambda}\frac{{\partial {R} }}{{\partial Q} } = 0,
\end{equation}
we have
\begin{equation}\label{formula:5}
{\lambda} =  - \frac{{\partial {D} }}{{\partial {R} }}.
\end{equation}

This implies that \(\lambda\) characterizes the \textit{slope} of the rate-distortion curve.
In our previous work \cite{xu2017}, we present the rate-distortion curves of different static point clouds, and show that the rate-distortion relationships of various static point clouds are quite close.
Therefore, we off-line derive the $\lambda$-Q model based on statistics of point clouds.
By discretizing the rate and distortion points in \eqref{formula:5}, we approximate the slope of the rate-distortion curve using neighboring rate and distortion points as
\begin{equation}\label{formula:6}
{\lambda_Q} =  - \frac{{{D_{Q + 1}} - {D_Q}}}{{{R_{Q + 1}} - {R_Q}}}.
\end{equation}

Given different dynamic point clouds, the relationships between $\lambda$ and quality factor $Q$ for these sequences are plotted in Fig.~\ref{fig55}. We acquire a best approximation of $\lambda$ with a power function of \(Q\) as
\begin{equation}\label{formula:8}
{\lambda _Q}{\rm{ = }}\alpha  \cdot {Q^\beta },
\end{equation}
where \(\alpha= 0.0624\) and \(\beta=1.6238\) via function fitting.

Based on the rate-distortion optimization and $\lambda$-Q model, the best mode for each cluster can be determined by choosing the one that leads to the smaller rate-distortion cost.

\section{Experimental Results}
\label{sec:results}
\begin{figure*}[t]
	\centerline{\psfig{figure=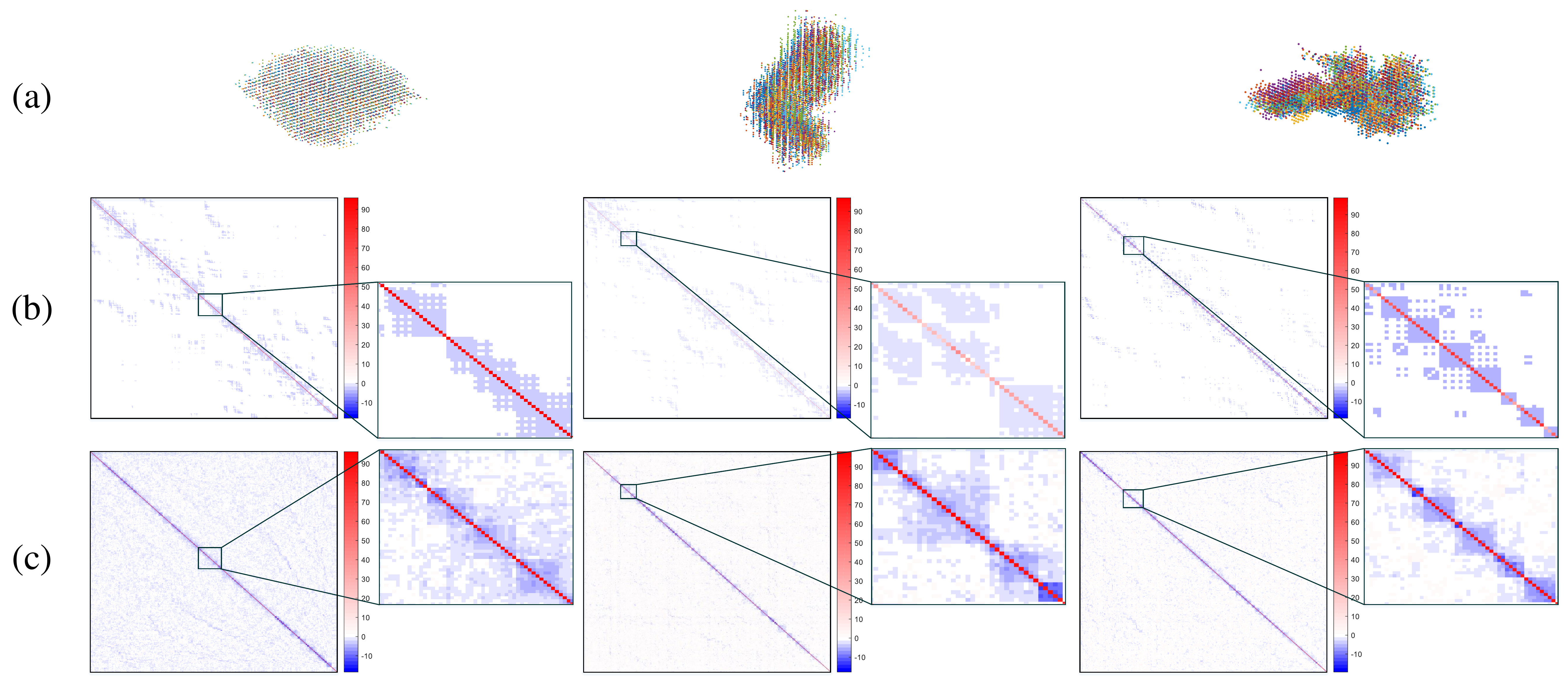,width=180mm} }
	\caption{Demonstration that the generalized graph Laplacian approximates the precision matrix of the data well. (a) shows three representative surface patches; for each surface patch, we have (b) the obtained generalized graph Laplacian with our graph construction method; and (c) the precision matrix computed from the data statistically. }
	\label{fig:LQ}
\end{figure*}

\begin{figure*}[htbp]

  \centering
  \centerline{\includegraphics[width=180mm]{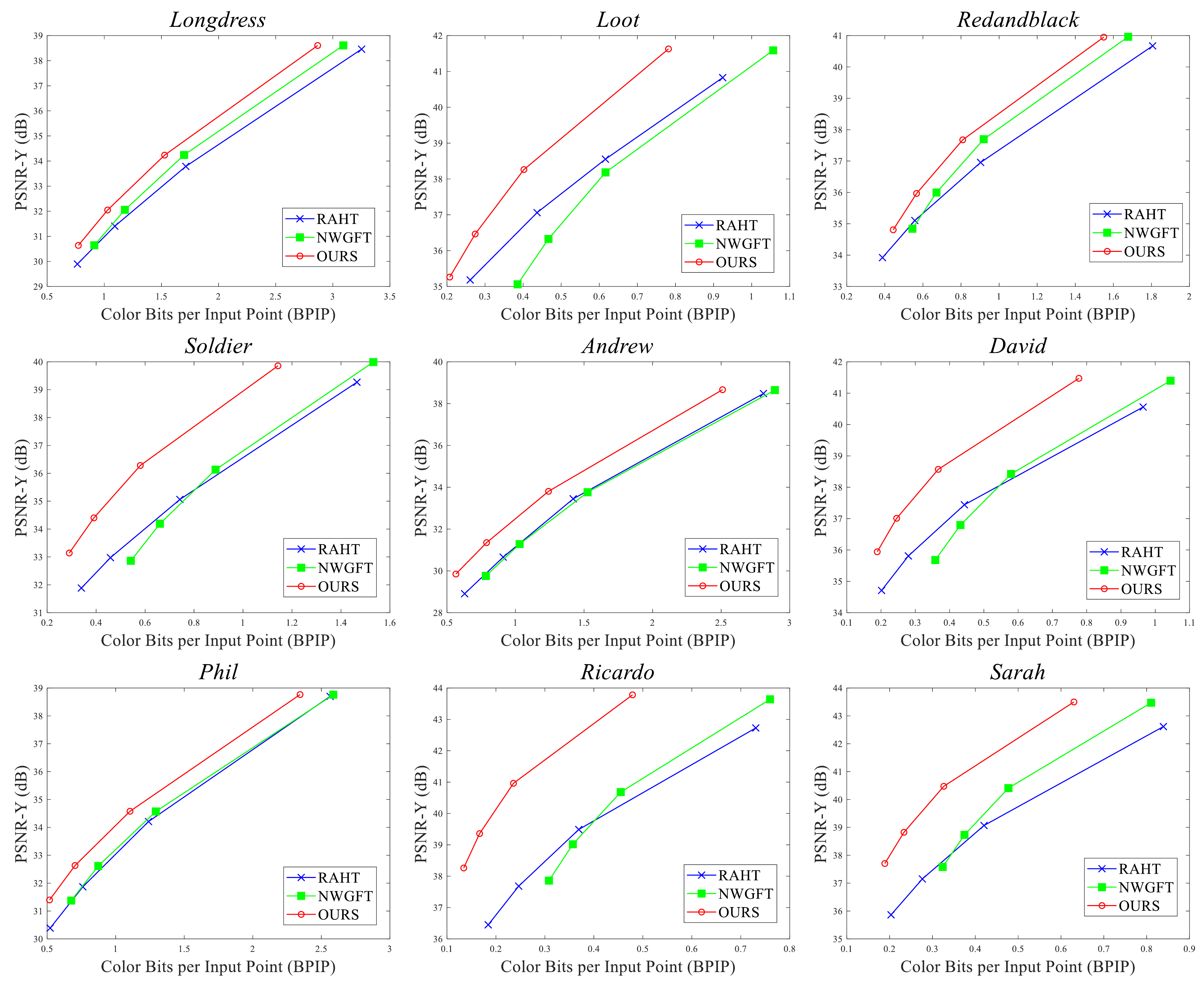}}
  \caption{Rate-Distortion curves for the proposed method, RAHT and NWGFT.}
  \label{fig:5}
%
\end{figure*}

\begin{figure*}[htbp]

  \centering
  \centerline{\includegraphics[width=180mm]{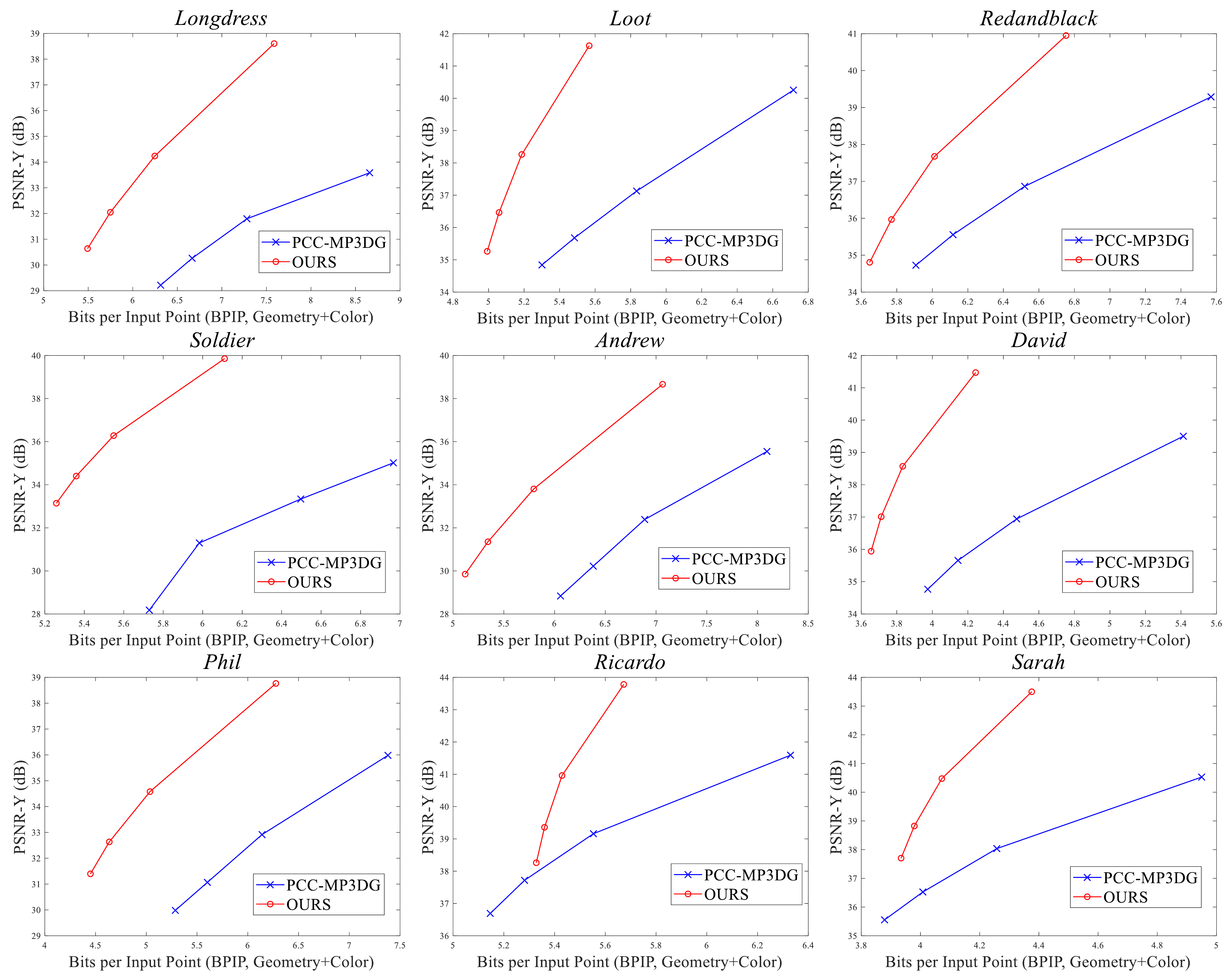}}
  \caption{Rate-Distortion curves for the proposed method and PCC-MP3DG.}
  \label{fig:rdcurve_mp3dg}
%
\end{figure*}

\begin{figure*}[htbp]

  \centering
  \centerline{\includegraphics[width=140mm]{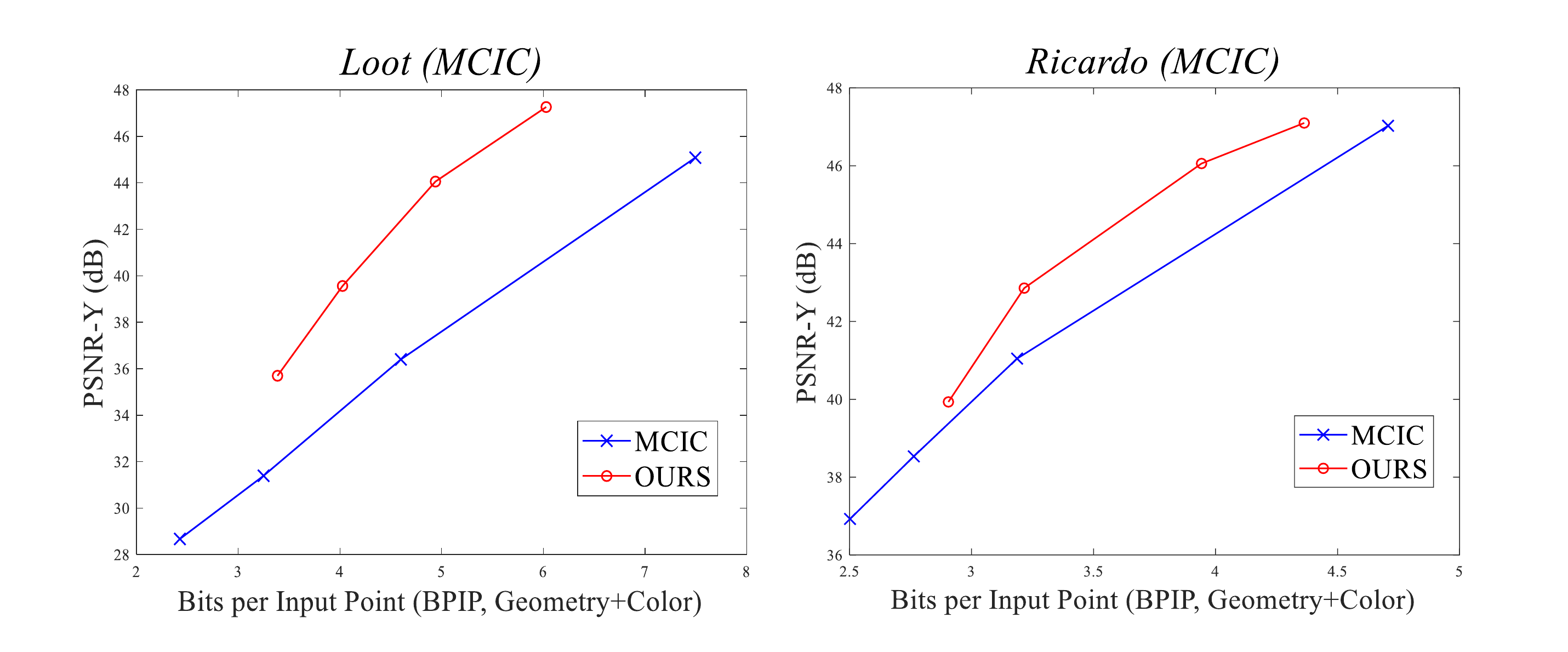}}
  \caption{Rate-Distortion curves for the proposed method and MCIC.}
  \label{fig:rdcurve_mcic}
%
\end{figure*}

\begin{figure*}[htbp]

  \centering
  \centerline{\includegraphics[width=180mm]{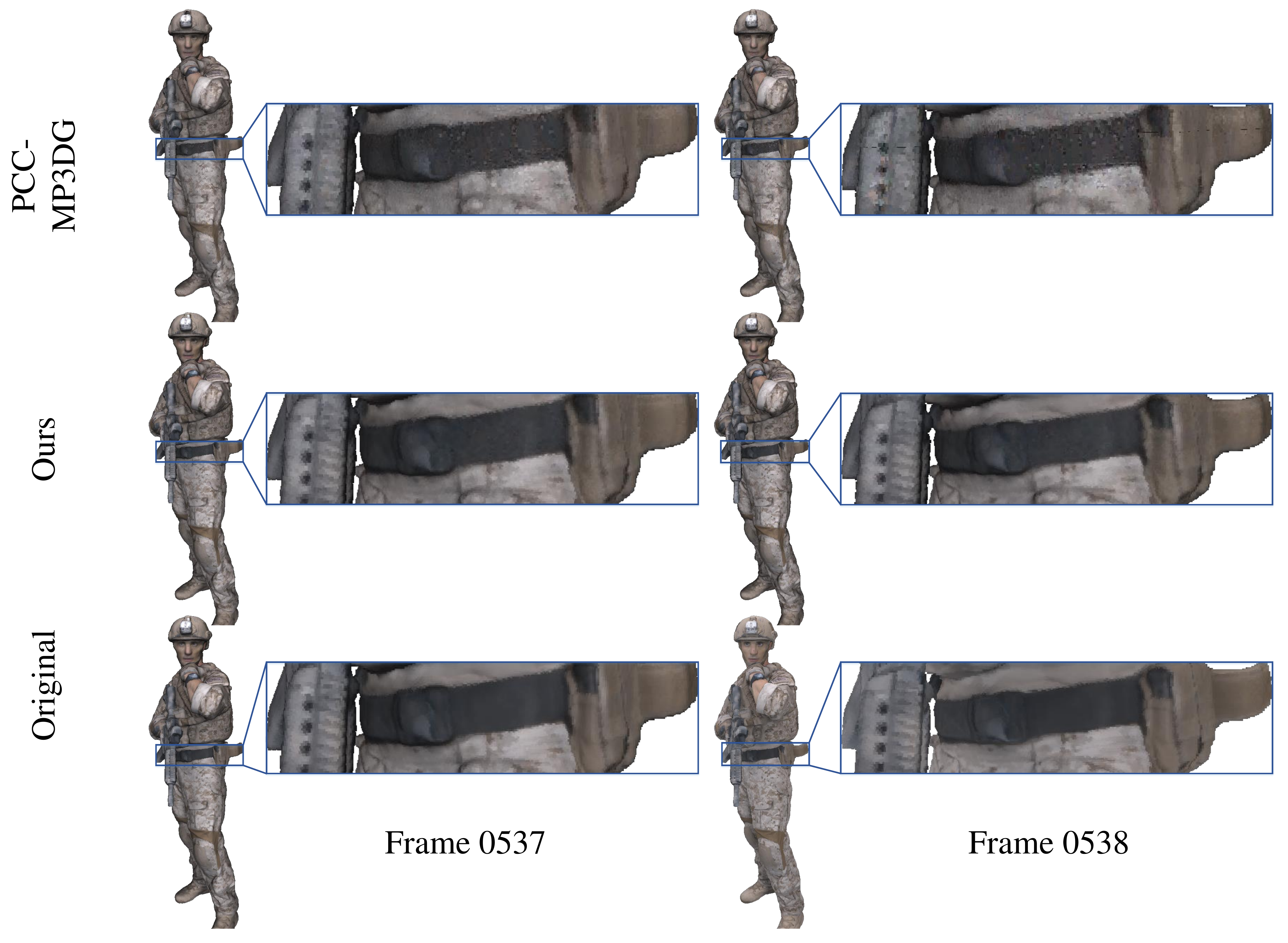}}
 \vspace{-0.1in}
  \caption{Subjective comparison with PCC-MP3DG on two reconstructed frames in the dynamic point cloud {\it Soldier}. They are encoded with similar rate on average: PCC-MP3DG: 6.13 BPIP, 31.96 dB for PSNR-Y; Ours: 6.11 BPIP, 39.86 dB for PSNR-Y.}
  \label{fig:sub_inter_mp3dg}
\end{figure*}

\begin{figure*}[htbp]

  \centering
  \centerline{\includegraphics[width=180mm]{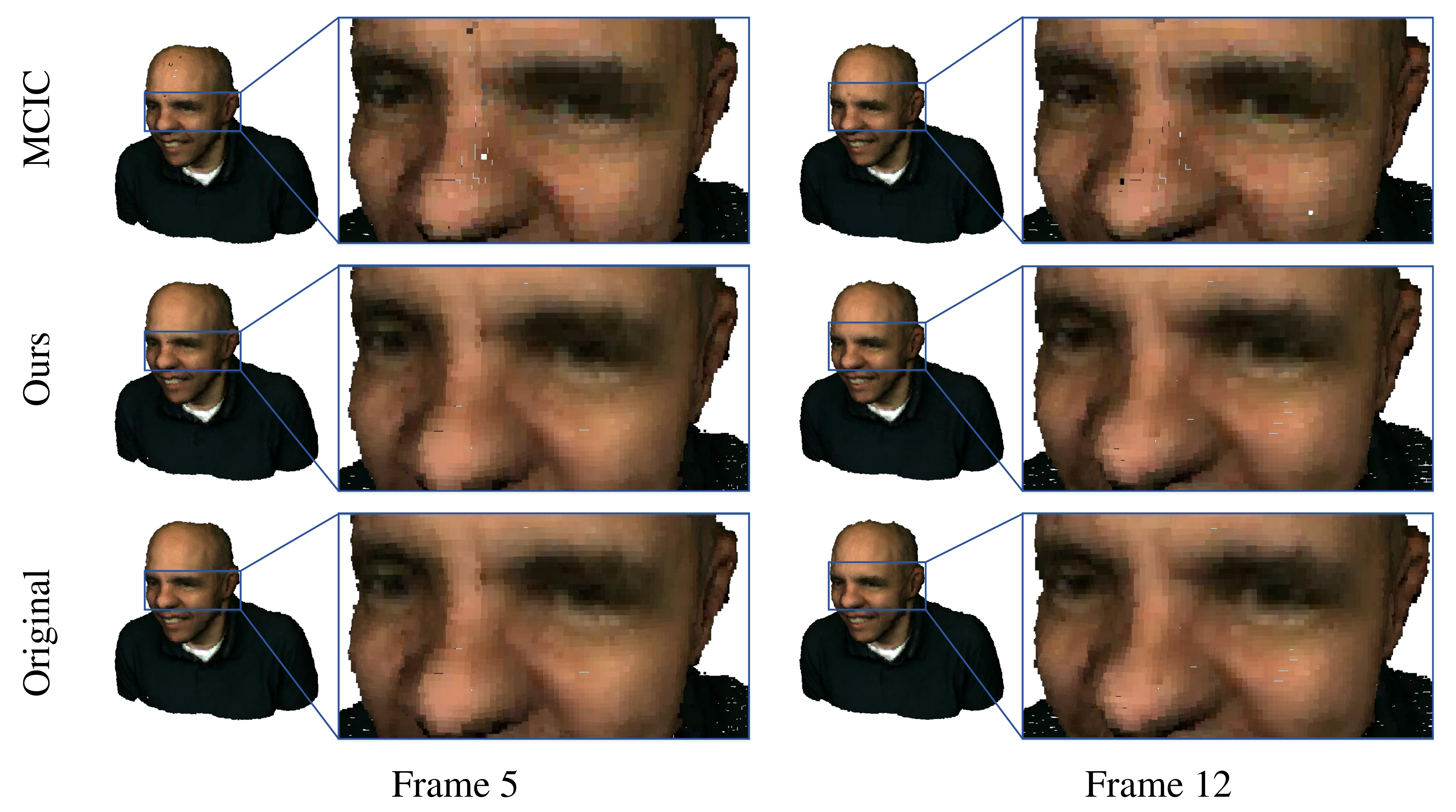}}
 \vspace{-0.1in}
  \caption{Subjective comparison with MCIC on two reconstructed frames in the dynamic point cloud {\it Ricardo (MCIC)}. They are encoded with similar rate on average: MCIC: 2.91 BPIP, 39.32 dB for PSNR-Y; Ours: 2.91 BPIP, 39.93 dB for PSNR-Y.}
  \label{fig:sub_inter_mcic}
\end{figure*}

\subsection{Experimental Setup}
To validate the proposed complete framework for attribute compression of dynamic 3D point clouds, we compare with two competitive static point cloud coding methods as well as two competitive {\it dynamic} point cloud coding methods with available codes:
1) Region-Adaptive Hierarchical Transform (RAHT) \footnote{https://github.com/digitalivp/RAHT.} \cite{de2016compression}, which has been adopted in geometry-based point cloud compression (G-PCC)\footnote{G-PCC is a widely deployed open-source point cloud compression API introduced by 3D Graphics (3DG) group of MPEG.} and is one state-of-the-art method for attribute coding of static point clouds \cite{3dg2019wd};
2) our previous intra-coding method Normal-Weighted Graph Fourier Transform (NWGFT) \cite{xu2018cluster};
3) PCC-MP3DG \cite{Mekuria2016}, which exploits both intra-frame and inter-frame dependencies and served as a base reference software platform in JTC1/SC29/WG11 (MPEG);
4) MCIC \cite{Ricardo2017mcic}, which proposes a novel motion-compensated approach to encoding dynamic voxelized point clouds at low bit rates.
All these methods are geometry-based as our approach for fair comparison.
As RAHT and NWGFT are proposed for \textit{static} point clouds, we perform them on each frame of a dynamic point cloud sequence independently for comparison.

We conduct experiments on two sources of datasets, including four MPEG sequences (\textit{Longdress}, \textit{Loot}, \textit{Redandblack} and \textit{Soldier}) from \cite{Eugene2017dataset} and five MSR sequences (\textit{Andrew9}, \textit{David9}, \textit{Phil9}, \textit{Ricardo9} and \textit{Sarah9}) from \cite{loop2016microsoft}.
For the convenience of experiments, 16 frames are selected from each sequence and the group of picture (GOP) is set to 8.
Low-delay P (LDP) configuration is adopted in our experiments.
We set the parameter $\epsilon^2$ in the $\epsilon$-neighborhood graph employed in Sec.~\ref{subsec:define_d} to $50$ for the first dataset and $300$ for the second dataset due to different densities of point clouds.
The reconstruction quality is calculated via the evaluation metric software for PCC in MPEG \cite{Tian2017eva}.
The Arithmetic coder \cite{rissanen1979arithmetic} is employed in entropy coding, and the bit rate is calculated via bit per input point (BPIP).

\begin{table}[htbp]
\renewcommand\arraystretch{1.15}
\centering
\caption{\label{tab:Performance2}Performance Comparison with NWGFT.}
\resizebox{80mm}{!}{
\begin{tabular}{c | c | c | c }
\hline
\hline

\bfseries{Point Clouds} & \bfseries{BD-BR (Y)} & \bfseries{BD-BR (U)} & \bfseries{BD-BR (V)}  \\

\hline
    \emph{$\textit{Longdress}$}              &  -10.2\%           &  -10.1\%   & -10.3\%    \\
    \emph{$\textit{Loot}$}                 &  -36.0\%           &  -43.9\%  & -44.0\%   \\
    \emph{$\textit{Redandblack}$}                &  -11.8\%           &  -10.7\%  & -12.0\%  \\
    \emph{$\textit{Soldier}$}               &  -36.1\%           &  -47.0\%  & -47.0\%   \\
    \emph{$\textit{Andrew}$}              &  -19.6\%           &  -25.2\%   & -25.3\%    \\
    \emph{$\textit{David}$}                 &  -38.4\%           &  -41.6\%  & -40.8\%   \\
    \emph{$\textit{Phil}$}                &  -14.8\%           &  -15.1\%  & -14.9\%  \\
    \emph{$\textit{Ricardo}$}               &  -50.2\%           &  -56.0\%  & -55.1\%   \\
    \emph{$\textit{Sarah}$}    &  -32.4\%           &  -34.9\%  & -34.0\%   \\

\hline
    \textbf{Average}                       &  -27.7\%           &  -31.6\%  & -31.5\%   \\
\hline
\hline
\end{tabular}
}
\end{table}

\begin{table}[htbp]
\renewcommand\arraystretch{1.15}
\centering
\caption{\label{tab:Performance1}Performance Comparison with RAHT.}
\resizebox{80mm}{!}{
\begin{tabular}{c | c | c | c }
\hline
\hline

\bfseries{Point Clouds} & \bfseries{BD-BR (Y)} & \bfseries{BD-BR (U)} & \bfseries{BD-BR (V)}  \\

\hline
    \emph{$\textit{Longdress}$}              &  -16.1\%           &  -28.0\%   & -24.8\%    \\
    \emph{$\textit{Loot}$}                 &  -28.6\%           &  -53.3\%  & -54.9\%   \\
    \emph{$\textit{Redandblack}$}                &  -21.3\%           &  -23.9\%  & -31.0\%  \\
    \emph{$\textit{Soldier}$}               &  -37.0\%           &  -57.6\%  & -58.4\%   \\
    \emph{$\textit{Andrew}$}              &  -17.5\%           &  -18.0\%   & -16.1\%    \\
    \emph{$\textit{David}$}                 &  -37.6\%           &  -39.8\%  & -35.2\%   \\
    \emph{$\textit{Phil}$}                &  -15.8\%           &  -15.4\%  & -14.0\%  \\
    \emph{$\textit{Ricardo}$}               &  -52.9\%           &  -60.8\%  & -59.9\%   \\
    \emph{$\textit{Sarah}$}    &  -40.8\%           &  -49.0\%  & -42.6\%   \\

\hline
    \textbf{Average}                       &  -29.7\%           &  -38.4\%  & -37.4\%   \\
\hline
\hline
\end{tabular}
}
\end{table}

\begin{table}[htbp]
\renewcommand\arraystretch{1.15}
\centering
\caption{\label{tab:mp3dg}Performance Comparison with PCC-MP3DG.}
\resizebox{40mm}{!}{
\begin{tabular}{c | c   }
\hline
\hline

\bfseries{Point Clouds} & \bfseries{BD-BR (Y)}  \\

\hline
    \emph{$\textit{Longdress}$}              &  -23.3\%                  \\
    \emph{$\textit{Loot}$}                 &  -14.1\%               \\
    \emph{$\textit{Redandblack}$}                &  -10.4\%               \\
    \emph{$\textit{Soldier}$}               &  -20.5\%                \\
    \emph{$\textit{Andrew}$}              &  -20.4\%                 \\
    \emph{$\textit{David}$}                 &  -20.4\%               \\
    \emph{$\textit{Phil}$}                &  -24.3\%               \\
    \emph{$\textit{Ricardo}$}               &  -6.7\%                \\
    \emph{$\textit{Sarah}$}    &  -11.6\%                \\

\hline
    \textbf{Average}                      &  -16.9\%              \\
\hline
\hline
\end{tabular}
}
\end{table}

\begin{table}[htbp]
\renewcommand\arraystretch{1.15}
\centering
\caption{\label{tab:mcic}Performance Comparison with MCIC.}
\resizebox{40mm}{!}{
\begin{tabular}{c | c   }
\hline
\hline

\bfseries{Point Clouds} & \bfseries{BD-BR (Y)}  \\

\hline
    \emph{$\textit{Loot (MCIC)}$}              &  -26.3\%                  \\
    \emph{$\textit{Ricardo (MCIC)}$}                 &  -8.3\%               \\

\hline
    \textbf{Average}                       &  -17.3\%              \\
\hline
\hline
\end{tabular}
}
\end{table}

\subsection{Evaluation of Precision Matrix Approximation}
\label{subsec:precision_approx}
In Section~\ref{sec:prediction}, we show that the proposed GGFT is optimal if the signal can be modeled by a GMRF with the precision matrix interpreted by a generalized graph Laplacian. We now validate this signal model by evaluating how well the generalized graph Laplacian obtained with the proposed graph construction method approximates the precision matrix of the data.

In particular, we choose three representative surface patches from {\it Ricardo}: a comparatively flat surface, a curved surface, and a surface with complicated structure, as shown in Fig.~\ref{fig:LQ}(a).
For each surface patch, we compute its generalized Laplacian matrix with the proposed graph construction method, whose entries are color coded as shown in Fig.~\ref{fig:LQ}(b).
Negative off-diagonal entries correspond to positive edge weights of the graph.
Then, we compute the precision matrix of each surface patch statistically from the dynamic point cloud of {\it Ricardo}.
Specifically, for each target surface patch in Fig.~\ref{fig:LQ}(a), we collect $K$ ($K=19$) most similar and aligned patches from the entire dynamic point cloud as in the refined motion estimation in Section IV-B, which are treated as multiple observations of the target surface patch. We then compute the covariance matrix from the $K+1$ surface samples, and take the inverse to acquire the precision matrix. 

Fig.~\ref{fig:LQ}(b) and Fig.~\ref{fig:LQ}(c) demonstrate the corresponding generalized graph Laplacian and precision matrix respectively.
For clearer comparison, we randomly magnify certain parts of both the generalized graph Laplacian and precision matrix for each surface patch, which describe a subset of the patch.
We observe that, on one hand, the generalized graph Laplacian shares similar prominent patterns with the precision matrix, {\it e.g.}, the structural patterns around the diagonal that describe prominent correlations among points.
On the other hand, the generalized graph Laplacian is generally sparser than the precision matrix.
This is because we build an $\epsilon$-neighborhood graph, which only connects points whose pairwise distances are smaller than $\epsilon$, while the precision matrix is dense as we compute the correlation between each pair of points.
In summary, this shows that the obtained generalized graph Laplacian is a good sparse approximation of the precision matrix statistically obtained from the data, thus validating our signal model.

\subsection{Experimental Results}

\subsubsection{Comparison with Static Point Cloud Coding Methods}

Fig.~\ref{fig:5} shows the rate-distortion curves for comparison with static point cloud coding methods---RAHT and NWGFT.
We see that our method significantly outperforms NWGFT and RAHT over all the test sequences for a large range of BPIP. Specifically, compared with NWGFT, we reduce the bit rate by 27.7\%, 31.6\% and 31.5\% on average for Y, U and V components respectively, as listed in Table~\ref{tab:Performance2}.
The numbers are calculated using the BD-BR \cite{Bjontegaard}, which quantifies the difference between two rate-distortion curves.
Also, comparing the measurement points on the rate-distortion curves, our method significantly reduces the bit rate with trivial loss in quality, due to our efficient inter-coding mode.

Compared with the state-of-the-art method RAHT, we reduce the bit rate by 29.7\%, 38.4\% and 37.4\% on average for Y, U and V components respectively, as presented in Table~\ref{tab:Performance1}.
In particular, for point clouds with slow motion and simple texture such as \textit{Ricardo} and \textit{Sarah}, our method reduces the bit rate by more than 40\% in the Y component.
Even for point clouds with richer texture information such as \textit{Andrew} and \textit{Phil}, our method achieves 17.5\% and 15.8\% of bit rate reduction in the Y component over RAHT respectively.

\subsubsection{Comparison with Dynamic Point Cloud Coding Methods}
We compare with representative dynamic point cloud coding methods---PCC-MP3DG and MCIC.
Since both of them jointly encode geometry and color, we integrate their geometry coding methods respectively into our attribute coding framework.
The geometry coding in PCC-MP3DG and MCIC differs, thus we compare with them separately as follows.

Firstly, to compare with PCC-MP3DG \cite{Mekuria2016}, we integrate the octree method for geometry coding in PCC-MP3DG into our attribute coding framework to acquire the bit rate of geometry.
Fig.~\ref{fig:rdcurve_mp3dg} shows the rate-distortion curves of PCC-MP3DG and our method.
We see that, our method significantly outperforms PCC-MP3DG on all the test datasets, which validates the effectiveness of our method. Specifically, we reduce the bit rate by 16.9\% on average over PCC-MP3DG, as listed in Table~\ref{tab:mp3dg}.
This verifies that the proposed optimal inter-prediction and GGFT lead to more compact representation for dynamic point clouds.

Further, we demonstrate the subjective comparison with PCC-MP3DG on {\it Soldier} under almost the same bit rate in Fig.~\ref{fig:sub_inter_mp3dg}.
As presented in the magnified region, the results of PCC-MP3DG exhibit blurred textures and artifacts in the smooth area.
In contrast, our results preserve the textures well and keep smoothness in the belt.  

Secondly, to compare with MCIC \cite{Ricardo2017mcic}, we integrate the octree method for geometry coding in MCIC into our attribute coding framework to acquire the bit rate of geometry.
As in MCIC, we test on {\it Loot (MCIC)} and {\it Ricardo (MCIC)}, which differ from the previously adopted {\it Loot} and {\it Ricardo} in terms of the sampling rate.
We cannot test on other datasets due to the lack of auxiliary motion vectors for MCIC or legal access to the {\it Man} sequence.
As shown in Fig.~\ref{fig:rdcurve_mcic}, our method outperforms MCIC by a large margin on both datasets.
Specifically, we reduce the bit rate by 17.3\% on average over MCIC, as listed in Table~\ref{tab:mcic}.
This again validates the effectiveness of our proposed method. 

Further, we demonstrate the subjective comparison with MCIC on {\it Ricardo (MCIC)} under the same bit rate in Fig.~\ref{fig:sub_inter_mcic}.
As presented in the magnified facial region, the results of MCIC exhibit severe blocking artifacts, while our results mitigate such artifacts and are close to the ground truth.


\section{Conclusion}
\label{sec:conclude}
We propose a complete compression framework for attributes of 3D dynamic point clouds, assuming availability of geometry at both the encoder and decoder. 
We represent dynamic point clouds on spatio-temporal graphs and model with Gaussian Markov Random Fields with respect to the underlying graphs, from which we derive optimal inter-prediction and predictive transforms as Generalized Graph Fourier Transforms for temporal decorrelation.
Also, we remove spatial redundancy by Normal-Weighted Graph Fourier Transforms in the intra-coding mode. 
The optimal coding mode is then determined based on rate-distortion optimization with the proposed offline-trained $\lambda$-Q model. 
Experimental results show that we significantly outperform competitive dynamic point cloud compression methods, thus validating the decorrelation effectiveness of the proposed framework.


\bibliographystyle{IEEEtran}
\bibliography{ref}

\end{document}